\documentclass[
aps,%
12pt,%
final,%
notitlepage,%
oneside,%
onecolumn,%
nobibnotes,%
nofootinbib,%
centertags]%
{revtex4}

\usepackage{graphics}
\usepackage{bm}
\usepackage[fleqn]{amsmath}
\usepackage{epsfig}
\usepackage{amssymb}

\newcommand{\be}{\begin{equation}}
\newcommand{\e}{\end{equation}}
\newcommand{\beml}{\begin{subequations}}
\newcommand{\eml}{\end{subequations}}
\newcommand{\beq}{\begin{eqnarray}}
\newcommand{\eq}{\end{eqnarray}}
\newcommand{\ba}{\begin{array}}
\newcommand{\ea}{\end{array}}
\newcommand{\unite}{\mathbf{\hat{e}}}
\newcommand{\lt}{\left}
\newcommand{\rt}{\right}
\newcommand{\n}{\nonumber}
\newcommand{\la}{\langle}
\newcommand{\ra}{\rangle}
\newcommand{\tr}{{\rm Tr}\,}

\newcommand{\re}{\,{\rm Re}\,}
\newcommand{\ep}{\boldsymbol{\varepsilon}}

\newcommand{\Du}{\textbf{D}^{\dagger}}
\newcommand{\Dd}{\textbf{D}}
\newcommand{\tD}{\overleftrightarrow{\boldsymbol{\Delta}}}
\newcommand{\vk}{\textbf{k}}

\newcommand{\bra}[1]{\left|#1\right\rangle}
\newcommand{\ket}[1]{\left\langle#1\right|}

\begin{document}

\date{\today}

\title{Spectrum of coherent backscattering of light by two atoms}

\author{\firstname{Vyacheslav} \surname{Shatokhin}}
\email{v.shatokhin@dragon.bas-net.by}
\affiliation{B.~I.~Stepanov
Institute of Physics, National Academy of Sciences, 220072  Minsk,
Belarus}

\begin{abstract}
We study theoretically inelastic spectrum of coherent
backscattering of laser light by two atoms.
For an intense laser field, there are frequency domains of
not only constructive but also destructive (self-)interference of the inelastic photons.
We interpret the emergent
spectral features using the dressed states and considering coherent backscattering
as a kind of the pump-probe
experiment.
\end{abstract}

\maketitle

\section{Introduction}

Remarkable progress in laser cooling and trapping of atomic gases
\cite{metcalf} that led to a realization of Bose-Einstein condensate
\cite{bec95} made it also possible an exploration of various
mesoscopic transport and localization phenomena \cite{meso94} using
cold atomic gases. In particular, possessing large (compared to
their geometric size) and easily manipulated scattering
cross-section, atoms turned out to be suitable for studying light
transport in the weak and, prospectively, strong localization
regimes. Recently, coherent backscattering (CBS) of resonant laser
light (an analog of weak localization of electrons in disordered
conductors) has been observed with cold, trapped atomic clouds
\cite{labeyrie99,kulatunga03}. Since then, CBS of light by cold
atoms has become an area of intense theoretical and experimental
research (for a recent review, see \cite{kupriyanov06}).

Coherent backscattering is an enhancement of the average intensity
of light reflected off a dilute, disordered medium in the
backscattering direction. The underlying physical reason for the
emergence of CBS is the constructive interference between the
counterpropagating (labelled ``direct'' and ``reversed'') multiple
scattering amplitudes. When a scattering medium consists of
individual atoms, several mechanisms affecting phase coherence
between the interfering amplitudes should be considered. These are
(i) Raman scattering on degenerate atomic transitions; (ii)
inelastic scattering; (iii) mechanical motion of atoms.

As regards the atomic degeneracy mediating Raman processes
accompanied by photon polarization flips, its dephasing role is
nowadays very well understood, with quantitative accordance between
theory \cite{mueller01,mueller02,kupriyanov03} and experiment
\cite{labeyrie03,kulatunga03}, for ensembles of Rb atoms.

The next dephasing mechanism, the inelastic photon scattering induced by
atomic saturation $s$, has been studied in less detail. Recent
experiment with cold Sr atoms demonstrated a
rapid decrease of CBS enhancement factor $\alpha$ versus saturation
at moderate $s\leq 1$ \cite{chaneliere03}. It was shown,
 within a scattering
theory approach applied to two atoms in the regime of weakly
nonlinear scattering \cite{wellens04}, that a decrease of CBS
enhancement factor occurs due to the partial distinguishability of
the interfering amplitudes. In the general case of many atoms, three
different amplitudes interfere constructively in the weakly
nonlinear regime, so that $\alpha$ may exceed the linear barrier $2$
\cite{wellens05}.

In this contribution, we will also be concerned with the impact of
saturation on CBS, though for arbitrary laser field intensities. But
prior to proceed with a presentation of this work, let us make a
brief note on the effect of thermal motion of atoms at temperatures on the order
100 $\mu$K typical for CBS experiments.

Mechanical motion spoils phase coherence between the direct and
reversed amplitudes \cite{golubentsev84} if spread $v$ of the atomic
velocities violates the resonance condition $kv\ll 2\gamma$
\cite{labeyrie00}, where $k$ is the wave number and $2\gamma$ is the
natural linewidth of the excited atomic state. In the regime of weak
laser intensities, this inequality is usually satisfied, and the
picture of motionless atoms works very well so long as the CBS
intensity is concerned \cite{mueller01}. However, already in the
elastic scattering regime, the photon recoil and Doppler effects do
modify the CBS spectrum \cite{kupriyanov04b}. It is even more so in
the inelastic scattering regime, when the atoms from the cloud are
rapidly accelerated out of resonance by a powerful laser field.
Nonetheless, we will ignore this acceleration and assume that the
atoms are fixed in space. Thus, we focus here on the laser field
coupling exclusively to the atomic internal degrees of freedom, in
order to highlight the fundamental interference effect under the
influence of the nonlinear scattering. Explanation of this influence
for atoms at rest is basic to its understanding for atoms in motion.

More specifically, we will study spectrum of CBS by two atoms in the
helicity preserving polarization channel. This topic is above all
motivated by our previous work \cite{shatokhin05,shatokhin06}, where
we established existense of the residual CBS contrast in the deep
saturation regime, due to the constructive (self-)interference of
inelastically scattered photons. However, this constructive
interference is a net effect of all inelastic photons. The question
that naturally arises, of what the character of interference is at a
given frequency, can only be answered after looking at the CBS
spectrum. In this work we answer this question for the particular
case of exact resonance. We demonstrate that, in the saturation
regime, there are frequency domains where the interferential
contribution exhibits not only constructive but also destructive
interference, and employ the pump-probe analysis and the dressed
states representation to identify the scattering processes that are
responsible for the emergent spectral features. Our results agree
with those derived within the Langevin equation approach
\cite{gremaud}.

The paper is organized as follows: We start with a brief presentation
of our model and the master equation approach that we are using. In Sect.~\ref{sect:results}
we present results for the stationary CBS intensity and enhancement factor, and thereafter for CBS
spectrum. In the last Section, we conclude our work.

\section{Master equation approach to CBS of light by two atoms}
\subsection{Model and the main quantity of interest}
\begin{figure}
\includegraphics[width=12cm]{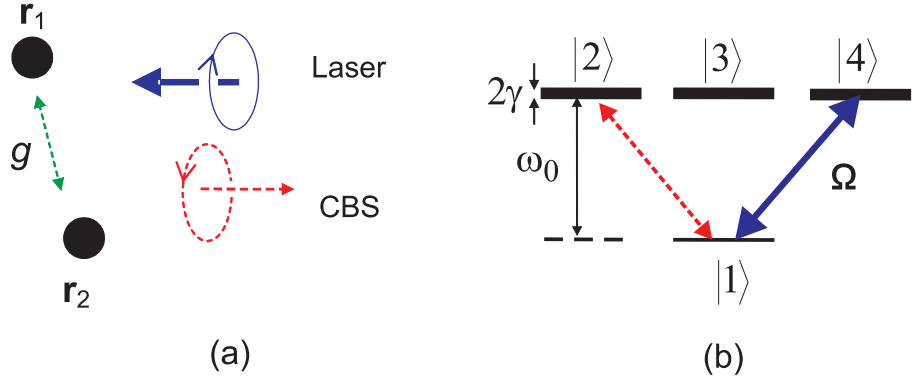}
\caption{ Model of CBS with two atoms. (a) atoms (black dots) are
driven by laser light with right circular polarization, while CBS is
observed in the helicity preserving channel, that is, with flipped
polarization. Photons in this channel appear as a result of double
scattering. $g$ is the strength of the far-field dipole-dipole
coupling responsible for exchange of photons; (b) internal atomic
structure corresponding to a $J_g=0\rightarrow J_e=1$ dipole
transition. $\omega_0$ is the transition frequency, $2\gamma$ is the
radiative linewidth, $\Omega$ is the Rabi frequency. Sublevels $\bra
1$ and $\bra 3$ have magnetic quantum number $m=0$. Sublevels $\bra
2$ and $\bra 4$ correspond to $m=-1$ and $m=1$, respectively. Thick
solid arrow shows laser field driving $|1\ra\leftrightarrow|4\ra$
transition, while dashed arrow shows CBS field originating from
$|1\ra\leftrightarrow|2\ra$ transition. } \label{fig:hph}
\end{figure}

Details of our approach are given in Ref.~\cite{shatokhin06}. Here, we will only present its brief
outline. We consider a model quantum system consisting of 2 identical, motionless atoms located
at positions ${\bf r}_1$ and ${\bf r}_2$, with the distance $r_{12}=|{\bf r}_1-{\bf r}_2|$
being much greater than the optical wavelength. The atoms are embedded in the electromagnetic
bath of quantized harmonic oscillators and subjected to an external laser
field of arbitrary intensity [See Fig.~\ref{fig:hph}(a)].
Coupling to the bath gives rise to the spontaneous emission from the
excited state and to the far-field dipole-dipole interaction
responsible for exchange of photons, whereas coupling to the laser
field gives rise to the Rabi oscillations of populations and
coherences in the laser-driven transitions of both atoms. Although
this approach can, of course, be formulated for atoms with arbitrary
internal structure, we choose the ground states of the atoms to be
nondegenerate, while the excited state 3-fold degenerate [see
Fig.~\ref{fig:hph}(b)]. An important parameter describing the effect
of a laser field on atoms is the so-called saturation parameter
$s=\Omega^2/2(\gamma^2+\delta^2)$, where $\Omega$ is the Rabi
frequency and $\delta=\omega_L-\omega_0$ is the detuning of the
laser field with respect to atomic resonance. As already mentioned,
here we will be interested in how the spectrum of CBS behaves as a
function of $s$ at $\delta=0$. In this case, all results can be
deduced in an analytic form.

Raman processes which can strongly affect CBS do not take place on the
$J_g=0\rightarrow J_e=1$ transition under consideration. Furthermore,
incoherent single scattering contribution can be filtered out by looking
at CBS, e.g., in the helicity preserving ($h\parallel h$) polarization channel. Precisely
this channel was probed in a recent experiment with cold Sr atoms \cite{chaneliere03}.
We consider the particular case of the laser light with the right circular polarization,
that is, $\ep_L=\unite_{+1}$, in the helicity basis notation. Hence, CBS with preserved helicity
corresponds to the flipped polarization
 $\ep=\unite_{-1}$ as shown on Fig.~\ref{fig:hph}.

Spectrum of CBS to be addressed in this paper is derived from the
average value of the first-order field temporal correlation function
\cite{glauber65}: \be G^{(1)}({\bf r},t;{\bf
r},t^\prime)=\lt\la\tr\{\rho[\ep\cdot{\bf E}^{(-)}({\bf r},t)]
[\ep^*\cdot{\bf E}^{(+)}({\bf r},t^\prime)]\}\rt\ra_{\rm conf},
\label{corr_func} \e where $\rho$ is the initial density operator of
the atom-field system, ${\bf E}^{(-/+)}({\bf r},t)$ is the
negative/positive frequency component of the electric field operator
of the scattered field, and $\lt\la\ldots\rt\ra_{\rm conf}$ denotes
configuration averaging. The components of the scattered field are
the retarded fields radiated by the atomic dipoles, \be {\bf
E}^{(+)}({\bf r},t) = \frac{\omega_0^2}{4\pi\varepsilon_0c^2r}
\sum_{\alpha=1}^2 \Dd_\alpha(t_\alpha) e^{-i\vk\cdot{\bf
r}_\alpha}\, , \label{Eplus} \e where $\epsilon_0$ is the
permittivity of the vacuum,
$\Dd_\alpha=-\unite_{-1}\sigma^\alpha_{12}+\unite_0\sigma^\alpha_{13}-\unite_{+1}\sigma^\alpha_{14}$,
with $\sigma^\alpha_{kl}\equiv\bra k_\alpha\ket l_\alpha$, is the
dipole lowering operator, and $t_\alpha=t-|{\bf r}-{\bf
r}_\alpha|/c$. In writing Eq.~(\ref{Eplus}), we have assumed that
$r_{12}\ll r$, that is, the field is detected in the radiation zone
at the distance much larger than the interatomic distance. In the
following, we will for brevity omit the $r$-dependent prefactor of
Eq.~(\ref{Eplus}) and, consistently, of the temporal correlation
functions.

Inserting Eq.~(\ref{Eplus}) into Eq.~(\ref{corr_func}) we obtain, in
the steady state limit $t\to\infty$, \be G^{(1)}_{\rm
ss}(\tau)=\sum_{\alpha,\beta=1}^2\la\la\sigma^\alpha_{21}\sigma^\beta_{12}(\tau)\ra_{\rm
ss} e^{i{\bf k}\cdot{\bf r}_{\alpha\beta}}\ra_{\rm conf},
\label{G_tau} \e where ``ss" stands for {\it steady state},
$\tau=t^{\prime}-t\geq 0$, the inner angular brackets indicate
quantum mechanical expectation value [see Eq.~(\ref{corr_func})],
and ${\bf r}_{\alpha\beta}\equiv {\bf r}_\alpha-{\bf r}_\beta$.

Spectrum can be obtained via Laplace
transform of (\ref{G_tau}) \cite{scully97}:
\be
S(\nu)=\frac{1}{\pi}\lim_{\Gamma\to 0}
\re\bigl\{\tilde{G}^{(1)}_{\rm ss}(z)\bigr\},
\label{sp_fixed}
\e
where $\tilde{G}^{(1)}_{\rm ss}(z)=\int_0^{\infty}d\tau\exp(-z\tau)G^{(1)}_{\rm ss}(\tau)$,
 $z=\Gamma-i\nu$ with $\Gamma\geq 0$ and $\nu=\omega-\omega_L$. Note that the
spectrum is defined with respect to the laser frequency which means that the
atomic correlation functions must be evaluated in the frame rotating at $\omega_L$.

Let us conclude this subsection with a remark on the configuration averaging procedure.
This procedure is necessary because the two-atom correlation functions
may sensitively depend on the interatomic distance and orientation of the vector ${\bf r}_{12}$ with
respect to ${\bf k}_L$, exhibiting rapid oscillations around the backscattering direction.
These oscillations have the same nature as a speckle pattern scattered off a disordered medium.
After many realizations of the disorder, all peaks except the one, corresponding to CBS, disappear.
A simple and sufficient way to mimic disorder in a two-atom system is to assume an
isotropic distribution
of the radius-vector connecting the atoms and a uniform distribution of
interatomic distances around the average distance $\ell$ equal to the scattering mean free path.

\subsection{Master equation}
To find the atomic correlation functions appearing in the right hand
side of Eq.~(\ref{G_tau}) we have adapted
\cite{shatokhin05,shatokhin06} a theoretical approach initiated by
Lehmberg in 1970 \cite{lehmberg70}. Within this approach, dynamics
of the dipole operators' expectation values as well as dipole-dipole
correlators is governed by the master equation \be \la\dot Q\ra
=\sum_{\alpha=1}^2\la{\cal L}_\alpha Q\ra+\sum_{\alpha\neq \beta=1}
^2\la{\cal L}_{\alpha\beta}Q\ra, \label{meq} \e where the
Liouvillians ${\cal L}_\alpha$ and ${\cal L}_{\alpha\beta}$ generate
the time evolution of an arbitrary atomic operator $Q$, for
independent and interacting atoms, respectively. Explicitly,
\begin{widetext}
\beq
{\cal L}_\alpha Q & = & -i\delta[\Du_\alpha\cdot\Dd_\alpha,Q]
-\frac{i}{2}[\Omega_\alpha(\Du_\alpha\cdot\ep_L)+\Omega^*_\alpha
(\Dd_\alpha\cdot\ep_L^*),Q]\nonumber\\
&&+\gamma\lt(\Du_\alpha\cdot[Q,\Dd_\alpha]+[\Du_\alpha,Q]\cdot\Dd_\alpha\rt),\label{L_a}
\\
{\cal L}_{\alpha\beta}Q&=&\Du_\alpha\cdot\overleftrightarrow{\bf
    T}(g,{\bf\hat
n})\cdot[Q,\Dd_\beta]+[\Du_\beta,Q]\cdot\overleftrightarrow{\bf
T}^*(g,{\bf\hat n})\cdot\Dd_\alpha\, ,
\label{L_ab}
\eq
\end{widetext}
where $\Omega_\alpha=\Omega e^{i{\bf k}_L\cdot{\bf r}_\alpha}$. The radiative
dipole-dipole interaction due to exchange of photons between the
atoms is described by the tensor $ \overleftrightarrow{\bf
T}(g,{\bf\hat n})=\gamma g
\overleftrightarrow{\boldsymbol{\Delta}}$, with
$\overleftrightarrow{\boldsymbol{\Delta}}=\overleftrightarrow
{\openone}-{\bf \hat{n}\hat{n}}$ being the projector on the
transverse plane defined by the unit vector $\bf\hat{n}$ along the
connecting line between the atoms $\alpha$ and $\beta$. This
interaction has a certain strength depending on the distance between
the atoms, via \be g =\frac{3i}{2k_0r_{\alpha\beta}}e^{ik_0
r_{\alpha\beta}}, \label{g} \e with $k_0=\omega_0/c$, and on the
life time of the excited atomic levels, through $\gamma$. The
coupling constant $|g|\ll 1$ is small in the far-field
($k_0r_{\alpha\beta}\gg 1$), where near-field interaction terms of
order $(k_0r_{\alpha\beta})^{-2}$ and $(k_0r_{\alpha\beta})^{-3}$
can be neglected.

Of course, an arbitrary operator $Q$ inserted into
Eq.~(\ref{meq}) does not result in a closed differential equation.
Our system consisting of two 4-level atoms leads to
$255=4^2\cdot 4^2-1$ linear coupled equations of motion for the one-time averages.
We solve them perturbatively up to $g^2$, to account for the lowest order (double-)scattering
process giving rise to a nontrivial interferential contribution. To help
the reader keeping this in mind we will supply symbols denoting double scattering intensities
and spectra with the subscript ``2''.

Note that Eq.~(\ref{meq}) describes evolution of the expectation
values (one-time correlation functions), whereas $G^{(1)}_{\rm
ss}(\tau)$ is the {\it two}-time correlation function. By virtue of
the quantum regression theorem \cite{scully97}, the latter satisfy
Eq.~(\ref{meq}) also, but their initial conditions are extracted
from the stationary solution of (\ref{meq}). In particular, the
double scattering counterpart of $G^{(1)}_{\rm ss}(0)$ is nothing
but the stationary average backscattered light intensity which will
be referred to as $I^{\rm tot}_2$. There is an obvious relation
between $I^{\rm tot}_2$ and $S_2(\nu)$: \be I^{\rm
tot}_2=\int_{-\infty}^{\infty}d\nu S_2(\nu). \label{I_vs_S} \e The
expression for $I^{\rm tot}_2$ can be obtained independently from
(\ref{I_vs_S}). We will use this independent derivation as an
implicit verification of our results for CBS spectra.

The total CBS intensity at the backscattering direction can be decomposed in the sum of two terms
\be
I^{\rm tot}_2=L^{\rm tot}_2+C^{\rm tot}_2,
\label{Itot2}
\e
where $C^{\rm tot}_2\equiv C^{\rm tot}_2(\theta=0)$ (i.e., ${\bf k}=-{\bf k}_L$), and
\beq
C^{\rm tot}_2(\theta)&=&2\re\la\la\sigma^1_{21}\sigma^2_{12}\ra^{[2]}_{\rm
  ss}e^{i\vk\cdot{\bf
r}_{12}}\ra_{\rm conf}\, ,\label{Cterm}\\
L^{\rm
tot}_2&=&\la\la\sigma_{22}^1\ra^{[2]}_{\rm
  ss}+\la\sigma_{22}^2\ra^{[2]}\ra_{\rm ss}\ra_{\rm
conf}\, ,\label{Lterm}
\eq
are the so called crossed and ladder terms, respectively.
Using these terms we can derive a standard measure of phase coherence
between the counterpropagating amplitudes in CBS -- the enhancement
factor
\be
\alpha=1+\frac{C^{\rm tot}_2}{L^{\rm tot}_2},
\label{efactor}
\e
which for perfect two-wave interference is equal to 2.

Generally, the total backscattered light intensity has the elastic
and inelastic counterparts, \be I^{\rm tot}_2=I^{\rm el}_2+I^{\rm
inel}_2. \e The elastic counterpart is given by the product of the
expectation values of the atomic dipoles, \be I^{\rm
el}_2=\sum_{\alpha,\beta=1}^2\la\la\sigma^\alpha_{21}\ra_{\rm
ss}\la\sigma^\beta_{12}\ra_{\rm ss} e^{-i{\bf k}_L\cdot{\bf
r}_{\alpha\beta}}\ra_{\rm conf}, \e wherefrom for $\alpha=\beta$ we
obtain the elastic ladder term $L^{\rm el}_2$ and for
$\alpha\neq\beta$ the elastic crossed term $C^{\rm el}_2$. Given
$I^{\rm tot}_2$ and $I^{\rm el}_2$ we can find the fluctuating part
of the dipole correlation functions defining $I^{\rm inel}_2$.

\section{Results}
\label{sect:results}
\subsection{CBS intensity and enhancement factor}
The interferential contribution and the incoherent sum, Eqs.~(\ref{Cterm}) and (\ref{Lterm}),
yield results \cite{shatokhin06}
\beq
2\re\{\la\sigma_{21}^1\sigma_{12}^2\ra_{\rm
ss}^{[2]}e^{i{\bf k}\cdot{\bf
r}_{12}}\}&=&|g|^2|\tD_{+1,+1}|^2\frac{R_1(s)}{(4+s)P(s)}\n\\
&&\times\cos\{({\bf k}+{\bf k}_L)\cdot{\bf r}_{12}\}\label{cohs},\\
\la\sigma_{22}^1\ra_{\rm ss}^{[2]}+\la\sigma_{22}^2\ra_{\rm
ss}^{[2]}&=&|g|^2|\tD_{+1,+1}|^2\frac{R_2(s)}{P(s)}\, . \label{pops}
\eq
$R_1(s)$, $R_2(s)$, and $P(s)$ are polynomial expressions in the
on-resonance saturation parameter $s=\Omega^2/2\gamma^2$,
\beml
\beq
R_1(s)&=&\frac{2}{9}\left(6912s+3168s^2\rt.\n\\
&&\lt.+264s^3+20s^4+s^5\rt),\\
R_2(s)&=&\frac{1}{3}\lt(1152s+528s^2+132s^3+7s^4\rt),\\
P(s)&=&(1+s)^2(12+s)(32+20s+s^2),
\eq
\label{RRP}
\eml
and $\tD_{+1,+1}=\unite_{+1}\cdot\tD\cdot\unite_{+1}$.

The configuration average of (\ref{cohs}) and (\ref{pops}) leads to
the final result \beq C^{\rm tot}_2(\theta)&\simeq&
\frac{|\tilde{g}|^2R_1(s)}{(4+s)P(s)}\Bigl(\frac{2}{15}-\frac{(k\,\ell\,\theta)^2}{35}\Bigr),
\label{CTheta}\\
L^{\rm tot}_2&=&\frac{2|\tilde{g}|^2R_2(s)}{15P(s)},\label{LHparH}
\eq
with $\tilde{g} = g|_{r_{\alpha\beta}=\ell}$. The scattering angle $\theta=2\arcsin\{|{\bf
k}+{\bf k}_L|/2k_L\}\ll 1$ with respect to the backscattering
direction was assumed to be sufficiently small herein.

\begin{figure}
\includegraphics[width=8cm]{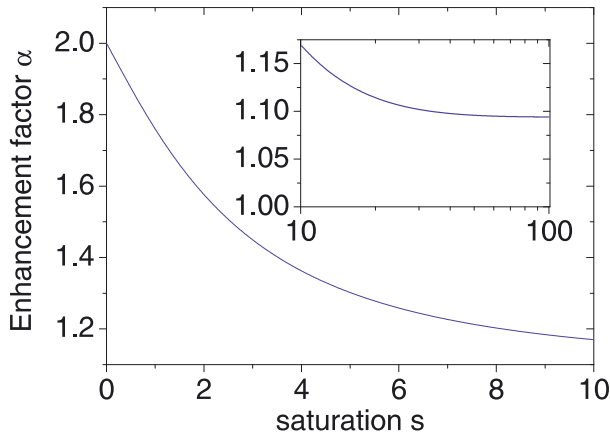}
\caption{ Enhancement factor in the $h\parallel h$ channel versus
saturation s. Decrease of $\alpha$ at small $s$ is described by the
linear function $2-s/4$ in accordance with \cite{wellens04}. Inset
describes $\alpha$ in the deep saturation regime. Enhancement tends
to the limit $\alpha_\infty=23/21$ \cite{shatokhin06} indicating
constructive self-interference of inelastic photons. }
\label{fig:enh}
\end{figure}

The enhancement factor $\alpha(s)$, Eq.~(\ref{efactor}), deduced from
Eqs.~(\ref{CTheta}) and (\ref{LHparH})
reads
\be
\alpha(s)=1+\frac{R_1(s)}{(4+s)R_2(s)}\, ,
\label{enh_res}
\e
and $\alpha(0)=2.0$ in the weak field limit, as
expected. The dependence of $\alpha$ on the saturation parameter is
shown on Fig.~\ref{fig:enh}.
For small $s$, enhancement linearly decreases as $2-s/4$, in full
agreement with the diagrammatic theoretical result \cite{wellens04}
and in qualitative agreement with the result of Sr experiment
\cite{chaneliere03}. When $s$ increases further, $\alpha$
monotonically drops to an asymptotic value
$\lim_{s\to\infty}\alpha(s)=\alpha_{\infty}=23/21$
\cite{shatokhin06} which is strictly larger than unity, implying a
nonvanishing residual CBS contrast in the limit of large injected
intensities. We will next show that this residual enhancement is due
to inelastic photons only. Indeed, we obtained the following result
for the elastic ladder and crossed terms \be L^{\rm el}_2=C^{\rm
el}_2=\frac{2|\tilde{g}|^2}{15}\frac{s}{(1+s)^4}. \label{LCel} \e As
seen from Eq.~(\ref{LCel}), the elastic component shows perfect
contrast for all $s$. In particular, it is this component that
results in enhancement $\alpha=2$ for very small $s\to 0$. However,
in the deep saturation regime, this component decreases as $s^{-3}$,
while the counterparts of the total intensity, Eqs.~(\ref{CTheta}),
(\ref{LHparH}), as $s^{-1}$. Herefrom follows our conclusion about
the origin of the residual enhancement in the deep saturation
regime. Explicitly, the inelastic crossed and ladder terms obtained
by elementary substraction of Eq.~(\ref{LCel}) from
Eqs.~(\ref{CTheta}) and (\ref{LHparH}) read \beq C^{\rm
inel}_2&=&\frac{2|\tilde{g}|^2}{15}\frac{20736s^2+23424s^3+7108s^4+601s^5+44s^6+2s^7}{9(1+s)^2(4+s)P(s)},
\label{Cinel}\\
L^{\rm
inel}_2&=&\frac{2|\tilde{g}|^2}{15}\frac{2016s^2+2244s^3+796s^4+146s^5+7s^6}{3(1+s)^2P(s)}.\label{Linel}
\eq It is easy to verify that $\lim_{s\to\infty}C^{\rm
inel}_2/L^{\rm inel}_2=2/21=\alpha_\infty-1$.

\subsection{CBS spectrum}
Double scattering spectrum of CBS has the elastic and inelastic
components. The elastic spectrum at the backscattering direction
reads \be \tilde{I}^{\rm el}_2(\nu)=I^{\rm el}_2\delta(\nu),
\label{Sp_el} \e where $\delta(\nu)$ is the Dirac's delta-function,
and $I^{\rm el}_2=L^{\rm el}_2+C^{\rm el}_2$, with the ladder and
crossed contributions defined in Eq.~(\ref{LCel}).

\begin{figure}
\includegraphics[width=14cm]{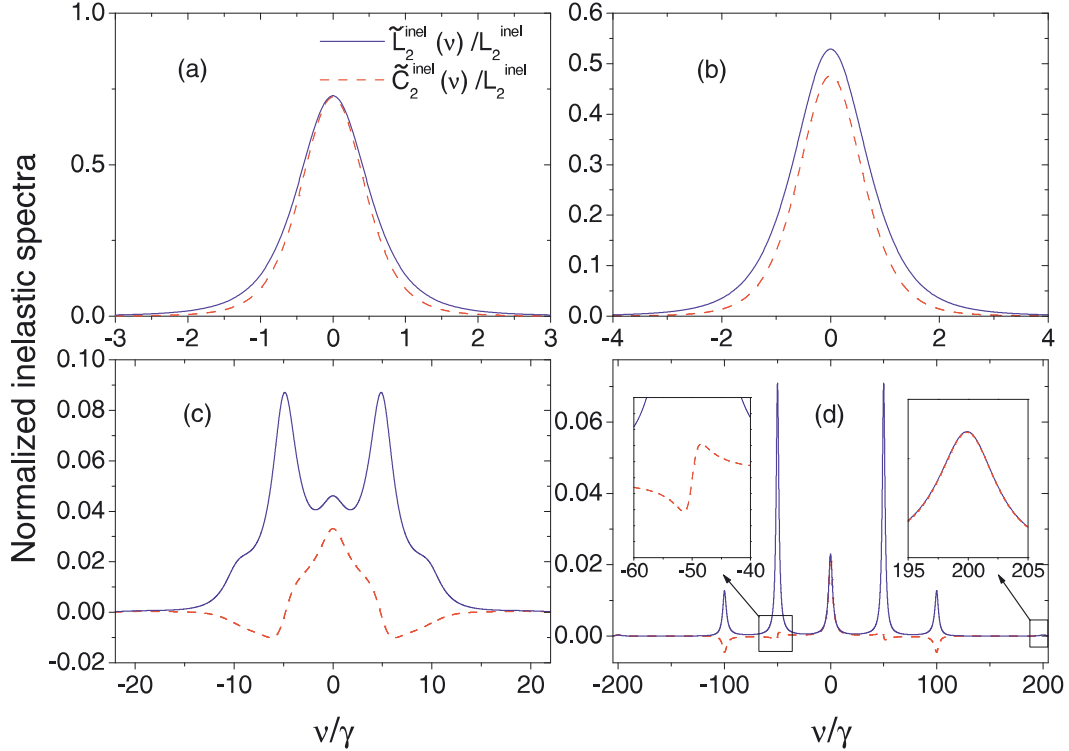}
\caption{Normalized inelastic spectra of the ladder (solid line) and
crossed (dashed line) terms at exact resonance for different values
of the Rabi frequency: (a) $\Omega=0.1\gamma$; (b) $\Omega=\gamma$;
(c) $\Omega=10\gamma$; (d) $\Omega=100\gamma$.}
\label{fig:spec_abcd}
\end{figure}

Inelastic spectra of the normalized ladder and crossed terms, for increasing values
of Rabi frequencies, are shown on Fig.~\ref{fig:spec_abcd}.
Normalization is chosen
such that integrals of $\tilde{L}^{\rm inel}_2(\nu)/L^{\rm inel}_2$ and
$\tilde{C}^{\rm inel}_2(\nu)/L^{\rm inel}_2$ over $\nu$ yield unity and
$C^{\rm inel}_2/L^{\rm inel}_2$, respectively. In the deep saturation
regime, the value of the latter integral tends to the asymptotic value of the interference
contrast of CBS, $\alpha_\infty-1$.

In describing the CBS spectrum, it is natural to use two parameters
$\Omega$ and $\gamma$ defining positions and linewidths of spectral
peaks rather than a single saturation parameter $s$. We will utilize
$s$ only to check consistency of our expressions for spectra with
the results for the inelastic intensity and the enhancement factor.

As seen from Fig.~\ref{fig:spec_abcd}, inelastic spectra are
symmetric with respect to the laser frequency, for all $\Omega$. For
a small value of the Rabi frequency $\Omega=0.1\gamma$
[Fig.~\ref{fig:spec_abcd}(a)], spectra of both the ladder and
crossed contributions have single peaks at $\nu=0$. Interferential
contribution $\tilde{C}_2^{\rm inel}(\nu)$ is positive, though
$\tilde{C}_2^{\rm inel}(\nu)\leq\tilde{L}^{\rm inel}_2(\nu)$
indicating that interference is not perfect in this weakly inelastic
regime. We can derive analytical expressions for the curves of
Fig.~\ref{fig:spec_abcd}(a) by leaving the leading-order
contribution to inelastic scattering $\sim (\Omega/\gamma)^4$,
corresponding to two-photon processes, and neglecting the
higher-order terms.

The ladder and crossed terms yield the compact expressions
(henceforth, we will omit the common prefactor $2|\tilde{g}|^2/15$):
\be \tilde{L}^{\rm
inel}_2(\nu)\simeq\frac{1}{\pi}\lt(\frac{\Omega}{\gamma}\rt)^4\frac{\gamma^3(2\gamma^2+\nu^2)}{2(\gamma^2+\nu^2)^3},
\quad \tilde{C}^{\rm
inel}_2(\nu)\simeq\frac{1}{\pi}\lt(\frac{\Omega}{\gamma}\rt)^4\frac{\gamma^5}{(\gamma^2+\nu^2)^3}.
\label{LCina} \e It is easy to establish that the expressions in
Eq.~(\ref{LCina}) are consistent with the behavior of the
enhancement factor in the two-photon scattering regime. Integrating
$\tilde{L}^{\rm inel}_2(\nu)$, $\tilde{C}^{\rm inel}_2(\nu)$ over
all frequencies, we obtain the following inelastic ladder and
crossed terms for small $\Omega$: \be L^{\rm
inel}_2\simeq\int^{\infty}_{-\infty} d\nu\tilde{L}^{\rm
inel}_2(\nu)=\frac{7}{16}\lt(\frac{\Omega}{\gamma}\rt)^4,\quad
C^{\rm inel}_2\simeq\int^{\infty}_{-\infty} d\nu\tilde{C}^{\rm
inel}_2(\nu)=\frac{3}{8}\lt(\frac{\Omega}{\gamma}\rt)^4.
\label{LandCin} \e Rewriting Eq.~(\ref{LandCin}) in terms of $s$ and
combining it with the small-$s$ expression for the elastic ladder
and crossed terms $L^{\rm el}_2=C^{\rm el}_2\simeq s$, we recover
the linear decrease \be \alpha=1+\frac{s+3s^2/2}{s+7s^2/4}\simeq
2-\frac{s}{4}. \e

As the Rabi frequency $\Omega$ increases further on [see
Fig.~\ref{fig:spec_abcd}(b,c,d)], qualitative differences emerge [on
Fig.~\ref{fig:spec_abcd}(c,d)] in the behavior of the ladder and
crossed terms. First, the spectra split into several distinct peaks.
Second, the crossed term becomes negative for a range of frequencies
beyond the central peak. This is a manifestation of destructive
self-interference of inelastically scattered photons. Note that a
similar effect of {\it antienhancement} was reported
\cite{kupriyanov04} for linear double scattering from atoms with
Zeeman-shifted hyperfine ground levels. New spectral features are
robust and become well-separated in the asymptotic limit of intense
driving [for an example at $\Omega=100\gamma$, see
Fig.~\ref{fig:spec_abcd}(d)]. We will next address the approximate analytic expression for
CBS spectrum at $\Omega\gg\gamma$, derived in the leading order $\sim (\gamma/\Omega)^2$.

In this case, the explicit expressions for the ladder and crossed
spectra can be represented by using a function of two real variables
$x_1$ and $x_2$: \be
\pounds(x_1,x_2)=\frac{1}{\pi}\frac{x_1}{x_1^2+x_2^2}.
\label{pounds} \e Let us mention the properties of
$\pounds(x_1,x_2)$ that are important to us: (i) if $x_1={\rm
Const}$, then function (\ref{pounds}) represents a Lorenzian with
width $x_1$ and a resonance at $x_2=0$; (ii) if $x_2={\rm Const}$,
then (\ref{pounds}) describes a resonance of the dispersive type at
$x_1=0$.

With the help of the function (\ref{pounds}), the ladder and crossed
spectra are given by
\beq \tilde{L}^{\rm
inel}_2(\nu)&\simeq&\lt(\frac{\gamma}{\Omega}\rt)^2\lt(\frac{1}{2}\pounds(\gamma,\nu)+
\frac{1}{4}\pounds(3\gamma,\nu)
+\frac{1}{72}[\pounds(3\gamma,\nu-2\Omega)+\pounds(3\gamma,\nu+2\Omega)]\rt.\n\\
&&\lt.+\frac{1}{9}[\pounds(3\gamma/2,\nu-\Omega)+\pounds(3\gamma/2,\nu+\Omega)]\rt.\n\\
&&\lt.+\frac{5}{18}[\pounds(5\gamma/2,\nu-\Omega)+\pounds(5\gamma/2,\nu+\Omega)]\rt.\n\\
&&\lt.+\frac{14}{9}[\pounds(3\gamma/2,\nu-\Omega/2)+\pounds(3\gamma/2,\nu+\Omega/2)]\rt),
\label{lad:asymp}\\ \tilde{C}^{\rm
inel}_2(\nu)&\simeq&\lt(\frac{\gamma}{\Omega}\rt)^2\lt(\frac{1}{2}\pounds(2\gamma,\nu)+
\frac{1}{4}\pounds(3\gamma,\nu)-\frac{1}{6}[\pounds(5\gamma/2,\nu-\Omega)+
\pounds(5\gamma/2,\nu+\Omega)]\rt.\n\\
&&\lt.+\frac{1}{72}[\pounds(3\gamma,\nu-2\Omega)+\pounds(3\gamma,\nu+2\Omega)]\rt)\n\\
&&+\lt(\frac{\gamma}{\Omega}\rt)^3\frac{208}{45}[\pounds(\nu+\Omega/2,3\gamma/2)
-\pounds(\nu-\Omega/2,3\gamma/2)], \label{cro:asymp} \eq where the
two terms of order $(\gamma/\Omega)^3$ are retained because they
define dispersive resonances of $\tilde{C}^{\rm inel}_2(\nu)$ at
$\nu=\pm\Omega/2$. As seen from Eqs.~(\ref{lad:asymp}) and
(\ref{cro:asymp}) as well as from Fig.~\ref{fig:spec_abcd}(d), both
the ladder and crossed terms have 7 resonances, the resonances of
both the ladder and crossed terms at $\nu=0$ and of the ladder term
at $\nu=\pm\Omega$ being sums of two Lorenzians with different
widths and weights $>0$. The rest resonances of the ladder term are
also Lorenzians with positive weights. Two resonances of the crossed
term at $\nu=\pm\Omega/2$ have the dispersive line shape and,
therefore, have no net contribution to the integrated intensity.
Furthermore, among the rest five resonances of the crossed term
which all are of the Lorenzian type, two at $\nu=\pm\Omega$ have
negative weights. Thus, inelastic photons (self-)interfere
destructively at $\nu=\pm\Omega$, yet the overall effect of all
inelastic processes is constructive. Note also that in the frequency
domains where interference is constructive, it is also perfect, as
can be concluded from the equality between the respective weights of
the ladder and crossed terms.

By performing the elementary integrations of Eqs.~(\ref{lad:asymp}) and
(\ref{cro:asymp}), we arrive at the inelastic ladder and crossed
terms \be L^{\rm inel}_2\simeq
\frac{14}{3}\lt(\frac{\gamma}{\Omega}\rt)^2,\quad C^{\rm
inel}_2\simeq \frac{4}{9}\lt(\frac{\gamma}{\Omega}\rt)^2,
\label{LCtot} \e which are consistent with Eqs.~(\ref{Cinel}), (\ref{Linel}) and, hence, with
$\alpha=\alpha_\infty=23/21$.

Let us now address the interpretation of the
CBS spectrum in the limit of intense driving.

\subsection{Interpretation}
One can understand the structure of the CBS spectrum from the
analysis of CBS as a specific realization of the pump-probe
experiment [a similar view is held by the authors of
Ref.~\cite{gremaud}]. In the usual setting of such an experiment
\cite{mollow72}, an atomic transition is simultaneously subjected to
two monochromatic fields: a variable-intensity, fixed-frequency
driving, or pump, field, and a weak probe field with tunable
frequency. For different frequencies of the probe field it can be
absorbed or amplified depending on the intensity of the pump field.
This occurs because the pump field leads to the energy levels'
shifts and broadenings, while the weak field transmission spectrum
probes these new resonances; hence the name of this technique.
\begin{figure}
\includegraphics[width=14cm]{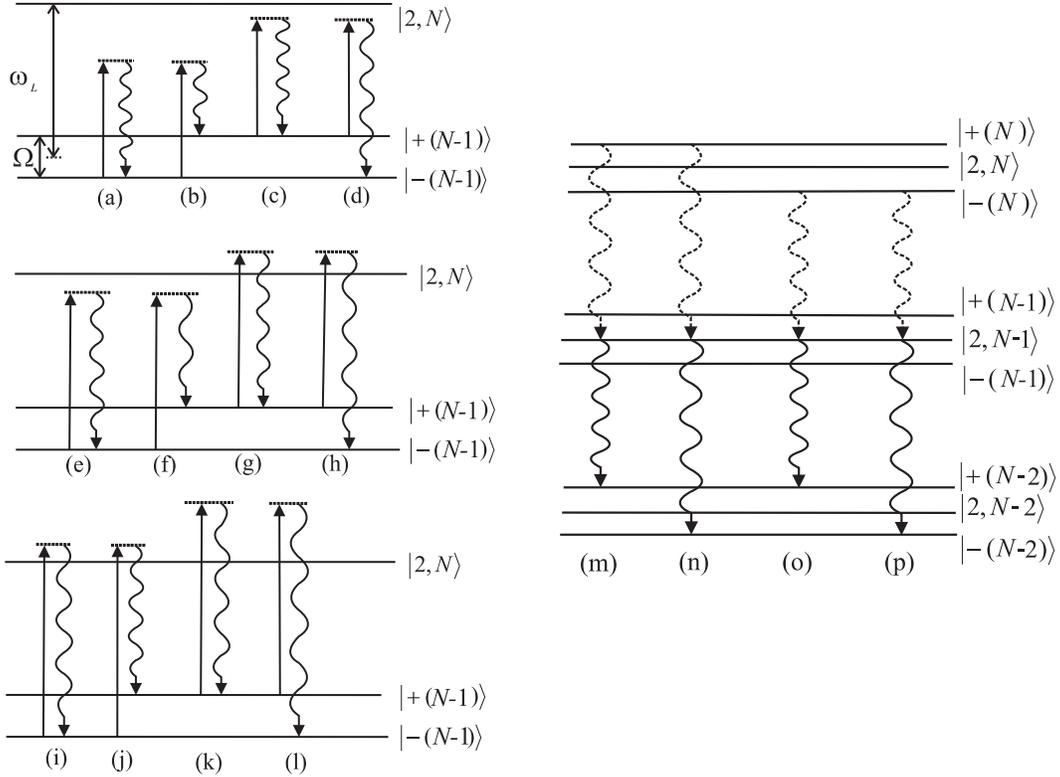}
\caption{
Scattering processes (a)-(l) and radiative transitions (m)-(p),
depicted by solid wavy arrows,
contributing to CBS spectra in the regime of intense driving.
Horisontal lines indicate dressed states.
In the processes (a)-(l), a photon with either of the frequencies
emitted by one atom ($\omega_L-\Omega$ (a-d); $\omega_L$ (e-h);
 $\omega_L+\Omega$ (i-l)) undergoes the Rayleigh or Raman
 scattering on the dressed states $|\pm(N-1)\ra$.
 Diagrams (m-n) show radiative cascades in which CBS resonances
 at $\omega_L\pm\Omega/2$ appear. Level $|2,N\ra$ can be populated as
 a result of a multiphoton scattering process with the participation of
one doubly scattered photon (depicted by dashed wavy arrows).
}
\label{fig:diags}
\end{figure}

In our case of CBS with two atoms, an intense pump acts in the
$|1\ra\leftrightarrow|4\ra$ transitions of both atoms, causing an AC
Stark shift of the energy levels. In this case it is instructive to
treat the laser mode as a quantum system strongly coupled to the
laser-driven atomic transition \cite{cohen_tannoudji}. The
eigenstates of the laser-atom interaction Hamiltonian for $\delta=0$
are the dressed states \be |\pm(N)\ra_\alpha=(|1,N+1\ra_\alpha\pm
e^{i{\bf k}\cdot{\bf r}_\alpha}|4,N\ra_\alpha)/\sqrt{2},
\label{dr_st} \e where $N$ and $N+1$ refer to the number of photons
in the laser mode, and $\alpha$ labels the atoms. Spontaneous
transitions from the dressed states manifold $\{|\pm(N)\ra_\alpha\}$
to $\{|\pm(N-1)\ra_\alpha\}$ lead to emission of the fluorescence
spectrum with three symmetric peaks centered at the frequencies
$\omega_L-\Omega$, $\omega_L$, and $\omega_L+\Omega$ known as the
Mollow triplet \cite{mollow69}. The Mollow triplet emitted by one
atom plays a role of the probe for another atom.

Figure \ref{fig:diags} illustrates the processes that contribute to
the CBS spectrum in $h \parallel h$ channel. The left part
Fig.~\ref{fig:diags}(a-l) shows possible one-photon elastic Rayleigh
and inelastic Raman processes which photons of frequencies
$\omega_L-\Omega$ (a-d), $\omega_L$ (e-h), and $\omega_L+\Omega$
(i-l) undergo on the dressed states $|\pm(N-1)\ra$. The right part
of Fig.~\ref{fig:diags}(m-n) shows radiative cascades in the dressed
state basis leading to resonances in the CBS spectrum at
$\omega_L\pm \Omega/2$. These lines appear as a result of the
spontaneous transitions from the state $|2,N-1\ra$ (note that the
atomic state $|2\ra$ is not affected by the laser field) to states
$|\pm(N-2)\ra$. This is the notable Autler-Townes doublet
\cite{cohen_tannoudji,autler55,mollow72b} emitted from the
transition between pairs of states of which only one is a member of
the laser-driven transition.

One observes that in all the processes except (b) and (l), several
transitions participate in the creation of a CBS photon. Phases
between the participating transitions can be opposite due to
difference, e.g., in the initial, intermediate, or final atomic
states, leading to the negative signs of the interferential
contributions around $\omega_L\pm\Omega$ and $\omega_L\pm\Omega/2$.

\section{Conclusion}
Using the master equation approach we have analytically calculated a
spectrum of CBS by two identical, motionless atoms for the case of
exact resonance between the laser and atomic transition frequencies.
This enabled us to analyse in more detail the previously established
\cite{shatokhin05,shatokhin06} effect of the constructive
self-interference of inelastically scattered photons.

One conclusion of this analysis is following. The enhancement factor
based on the total backscattered light intensity is a poor measure
of phase coherence between the counterpropagating waves in the
saturation regime, because it has the important information on the
character of interference at a given frequency integrated out. At
intense driving, one should rather use spectrally resolved
measurements with a filter whose passband $\Gamma_{\rm f}$ satisfies
$\gamma\ll\Gamma_{\rm f}\ll\Omega$. Then, tuning the filter on
individual peaks of the CBS spectra, one would observe either
perfect enhancement (at $\omega=\omega_L; \omega_L\pm 2\Omega$), or
antienhacement (at $\omega=\omega_L\pm\Omega$), else no net
interference at all (at $\omega=\omega_L\pm\Omega/2$).

Another conclusion is that spectral features of CBS can be
qualitatively understood from analysis of CBS as a kind of the
pump-probe experiment. In the limit of an intense driving, when the
spectral line-shape looks rather complicated [see
Fig.~\ref{fig:diags}], the `pump-probe' interpretation in
combination with the dressed states approach allowed us to identify
the origin of all the resonances of the CBS spectrum. Although a
detailed explanation of the character of interference between the
different processes, in which photons from the probe field
(scattered by one atom) are scattered on the dressed states (of
another atom), is beyond the scope of this work, the following
remark is in order. At exact resonance, some of the processes
interfere constructively while some destructively, with the overall
effect being constructive. But it is possible to vary populations of
the dressed states and, consequently, the weights of different
processes by changing the laser-atom detuning $\delta$. Nothing
forbids the overall effect of inelastic photons to be destructive.
In fact, we showed in the previous work \cite{shatokhin06} that for
large detuning it is indeed so. We also established in the same
paper that the ladder term of double scattering becomes negative in
$h\perp h$ channel in the saturation regime. This result can be
interpreted as a mere absorption of the probe field.

A challenging problem for the future would be to generalize the
results of the present work for moving atoms.

\begin{acknowledgements}
I would like to thank Andreas Buchleitner and Cord M\"uller for the
fruitful collaboration and encouragement which made this work
possible. Useful discussions with D.~Delande, B.~Gr\'emaud,
S.~Ya.~Kilin, D.~V.~Kupriyanov, C.~Miniatura, A.~P.~Nizovtsev,
M.~O.~Scully, I.~M.~Sokolov, M.~Titov, C.~Viviescas, and T.~Wellens
are gratefully acknowledged. Last not least, it is a pleasure to
thank the members of the Research Group ``Nonlinear Dynamics in
Quantum Systems'' at the Max Planck Institute for the Physics of
Complex Systems for the friendly and creative atmosphere during my
visits there.

\end{acknowledgements}

\end{document}